\documentstyle[12pt,epsf]{article}
\def\fo{\hbox{{1}\kern-.25em\hbox{l}}}

\newcommand{\newc}{\newcommand}

\newc{\lcal}{\int {\cal L}dt}

\newc{\LSP}{{\chi^0_1}}
\newc{\stauR}{{\tilde \tau_R}}
\newc{\stau}{{\tilde \tau_1}}
\newc{\mstop}{m_{\tilde{t}}}
\newc{\mHpm}{m_{H^\pm}}
\newc{\gsim}{\lower.7ex\hbox{$\;\stackrel{\textstyle>}{\sim}\;$}}
\newc{\lsim}{\lower.7ex\hbox{$\;\stackrel{\textstyle<}{\sim}\;$}}
\newc{\ie}{{\it i.e.}}
\newc{\etal}{{\it et al.}}
\newc{\eg}{{\it e.g.}}
\newc{\kev}{\hbox{\rm\,keV}}
\newc{\mev}{\hbox{\rm\,MeV}}
\newc{\gev}{\hbox{\rm\,GeV}}
\newc{\tev}{\hbox{\rm\,TeV}}
\newc{\xpb}{\hbox{\rm\, pb}}
\newc{\xfb}{\hbox{\rm\, fb}}

\newc{\mtop}{m_t}
\newc{\mbot}{m_b}
\newc{\mz}{m_Z}
\newc{\mw}{M_W}
\newc{\alphasmz}{\alpha_s(m_Z^2)}
\newc{\swsq}{\sin^2\theta_W}
\newc{\tw}{\tan\theta_W}
\newc{\cw}{\cos\theta_W}
\newc{\sw}{\sin\theta_W}
\newc{\BR}{\hbox{\rm BR}}
\newc{\zbb}{Z\to b\bar}
\newc{\Gb}{\Gamma (Z\to b\bar b)}
\newc{\Gh}{\Gamma (Z\to \hbox{\rm hadrons})}
\newc{\rbsm}{R_b^\hbox{\rm sm}}
\newc{\rbsusy}{R_b^\hbox{\rm susy}}
\newc{\drb}{\delta R_b}

\newc{\sgn}{\mbox{sgn}}

\newc{\tbeta}{\tan\beta}
\newc{\uL}{{\tilde u_L}}
\newc{\uR}{{\tilde u_R}}
\newc{\cL}{{\tilde c_L}}
\newc{\cR}{{\tilde c_R}}
\newc{\tL}{{\tilde t_L}}
\newc{\tR}{{\tilde t_R}}
\newc{\dL}{{\tilde d_L}}
\newc{\dR}{{\tilde d_R}}
\newc{\sL}{{\tilde s_L}}
\newc{\sR}{{\tilde s_R}}
\newc{\bL}{{\tilde b_L}}
\newc{\bR}{{\tilde b_R}}
\newc{\eL}{{\tilde e_L}}
\newc{\eR}{{\tilde e_R}}
\newc{\mhp}{m_{H^\pm}}
\newc{\mhalf}{m_{1/2}}
\newc{\emt}{{e/\mu /\tau}}

\newc{\lR}{\tilde{l}_R}
\newc{\lL}{\tilde{l}_L}
\newc{\nL}{\tilde{\nu}_L}
\newc{\na}{\chi^0_1}
\newc{\nb}{\chi^0_2}
\newc{\nc}{\chi^0_3}
\newc{\nd}{\chi^0_4}
\newc{\ca}{\chi^{\pm}_1}
\newc{\cb}{\chi^{\pm}_2}
\newc{\camp}{\chi^\mp_1}
\newc{\cbmp}{\chi^\mp_1}
\newc{\capos}{\chi^{+}_1}
\newc{\caneg}{\chi^{-}_1}
\newc{\phit}{\phi_t}
\newc{\phib}{\phi_b}
\newc{\phiew}{\phi_{ew}}
\newc{\htz}{h^0_t}
\newc{\hbz}{h^0_b}
\newc{\hewz}{h^0_{ew}}
\newc{\hsmz}{h^0_{sm}}
\newc{\huz}{h^0_u}
\newc{\hsusyz}{h^0_{susy}}

\def\mp{M_P}

\def\beq{\begin{equation}}
\def\eeq{\end{equation}}
\def\bea{\begin{eqnarray}}
\def\eea{\end{eqnarray}}
\def\slashchar#1{\setbox0=\hbox{$#1$}           
   \dimen0=\wd0                                 
   \setbox1=\hbox{/} \dimen1=\wd1               
   \ifdim\dimen0>\dimen1                        
      \rlap{\hbox to \dimen0{\hfil/\hfil}}      
      #1                                        
   \else                                        
      \rlap{\hbox to \dimen1{\hfil$#1$\hfil}}   
      /                                         
   \fi}                                         %
\catcode`@=11
\long\def\@caption#1[#2]#3{\par\addcontentsline{\csname
  ext@#1\endcsname}{#1}{\protect\numberline{\csname
  the#1\endcsname}{\ignorespaces #2}}\begingroup
    \small
    \@parboxrestore
    \@makecaption{\csname fnum@#1\endcsname}{\ignorespaces #3}\par
  \endgroup}
\catcode`@=12

\def\jfig#1#2#3{
 \begin{figure}
 \centering
 \epsfysize=4.6in
 \epsffile{#2}
 \caption{#3}
 \label{#1}
 \end{figure}}

\begin{document}

\baselineskip=18pt

\begin{titlepage}
\begin{flushright}
IZTECH-P/2005-02
\end{flushright}

\begin{center}
\vspace{1cm}

{\Large \bf Higher Curvature Quantum Gravity and Large Extra Dimensions}

\vspace{0.5cm}

{\bf Durmu{\c s} A. Demir and {\c S}{\"u}kr{\"u} H. Tany{\i}ld{\i}z{\i}}

\vspace{.8cm}

{\it Department of Physics, Izmir Institute of Technology, IZTECH, TR35430, Izmir, Turkey}

\end{center}
\vspace{1cm}

\begin{abstract}
\medskip
We discuss effective interactions among brane matter induced by modifications of higher dimensional Einstein
gravity via the replacement of Einstein-Hilbert term  with a generic function $f({\cal{R}})$ of the curvature
scalar ${\cal{R}}$. After deriving the graviton propagator, we analyze impact of virtual graviton exchanges on
particle interactions, and conclude that $f({\cal{R}})$ gravity effects are best probed by high-energy processes
involving massive gauge bosons, heavy fermions or the Higgs boson. We perform a comparative analysis of the
predictions of $f({\cal{R}})$ gravity and of Arkani-Hamed--Dvali--Dimopoulos (ADD) scenario, and find that the
former competes with the latter when $f^{\prime\prime}(0)$ is positive and comparable to the fundamental scale of
gravity in higher dimensions. In addition, we briefly discuss graviton emission from the brane as well as its
decays into brane-localized matter, and find that they hardly compete with the ADD expectations. Possible
existence of higher-curvature gravitational interactions in large extra spatial dimensions opens up various
signatures to be confronted with existing and future collider experiments.

\end{abstract}

\bigskip
\bigskip

\begin{flushleft}
IZTECH-P/2005-02 \\
November 2005
\end{flushleft}

\end{titlepage}


\section{Introduction}

The relative feebleness of gravity with respect to the weak force and its stability under quantum fluctuations,
the gauge hierarchy problem, has been pivotal for introducing a number of 'new physics' models to complete the
standard electroweak theory (SM) above Fermi energies. The idea \cite{idea3,idea4,idea5} that the scale of
quantum gravity can be much lower than the Planck scale, possibly as low as the electroweak scale itself
\cite{idea2,idea1} (see also the recent standard-like models found in intersecting D-brane models \cite{ibanez})
since this extreme is not excluded by the present experimental bounds \cite{exp}, has opened up novel lines of
thought and a number of phenomena which possess observable signatures in laboratory, astrophysical and
cosmological environments.

The basic setup of the Arkani-Hamed--Dimopoulos--Dvali (ADD) scenario \cite{idea1} is that $(1+3)$--dimensional
universe we live in is a field-theoretic brane \cite{rubakov} which traps all flavors of matter except the SM
singlets $e.g.$ the  graviton and right-handed neutrinos. As long as the surface tension of the brane does not
exceed the fundamental scale $\overline{M}_D$ of $D$--dimensional gravity, at distances $\gg 1/\overline{M}_D$
the spacetime metric $g_{A B}$ remains essentially flat. In other words, for singlet emissions (from brane) with
transverse (to brane) momenta $\left|\vec{p}_T\right| \ll \overline{M}_D$ the background spacetime is basically
Minkowski. Therefore, it is admissible to expand $D$--dimensional metric about a flat background
\begin{eqnarray}
\label{metric} g_{A B} = \eta_{A B} + 2 \overline{M}_D^{1-D/2} h_{A B}
\end{eqnarray}
where $\eta_{A B} =\mbox{diag.}\left(1, -1, -1, \cdots, -1\right)$ and $h_{A B}$ are perturbations. The
gravitational sector is described by Einstein gravity
\begin{eqnarray}
\label{ah} S_{ADD}=\int d^{D}x\, \sqrt{-g} \left\{ - \frac{1}{2} \overline{M}_D^{D-2} {\cal{R}} +
{\cal{L}}_{matter}\left(g_{A B}, \psi\right)\right\}
\end{eqnarray}
where $\psi$ collectively denotes the matter fields localized on the brane. There are various ways \cite{idea1}
to see that the Planck scale seen on the brane is related to the fundamental scale of gravity in higher
dimensions via
\begin{eqnarray}
\label{planck} \overline{M}_{Pl} = \sqrt{V_{\delta}} \overline{M}_D^{1+\delta/2}
\end{eqnarray}
which equals $(2 \pi R)^{1/2} \overline{M}_D^{1+\delta/2}$ when $\delta\equiv D-4$ extra spatial dimensions are
compactified over a torus of radius $R$. Obviously, larger the $R$ closer the $\overline{M}_D$ to the electroweak
scale \cite{idea1}. Experimentally, size of the extra dimensions, $R$, can be as large as a small fraction of
millimeter \cite{exp}, and thus, quantum gravitational effects can already show up at experimentally accessible
energy domains provided that the strength of gravitational interactions on the brane drives from higher
dimensional gravity as in (\ref{planck}). Upon compactification, the higher dimensional graviton gives rise to a
tower of massive S, P and D states on the brane, and they participate in various scattering processes involving
radiative corrections to SM parameters, missing energy signals as well as graviton exchange processes. These
processes and their collider signatures have been discussed in detail in seminal papers \cite{giudice,han}.

The ADD mechanism is based on higher dimensional Einstein gravity with metric (\ref{metric}). Given the very fact
that general covariance does not forbid the action density in (\ref{ah}) to be generalized to a generic function
$f\Big({\cal{R}},$ $\Box {\cal{R}},$ $\nabla_{A} {\cal{R}} \nabla^{A} {\cal{R}},$ ${\cal{R}}_{A B} {\cal{R}}^{A
B},$ ${\cal{R}}_{A B C D} {\cal{R}}^{A B C D}, \dots\Big)$ of curvature invariants, in this work {\it we will
derive and analyze effective interactions among brane matter induced by such modifications of higher dimensional
Einstein gravity, and compare them in strength and structure with those predicted by the ADD mechanism.} The
simplest generalization of (\ref{ah}) would be to consider, as we will do in what follows, a generic function
$f({\cal{R}})$ of the curvature scalar. Such modified gravity theories are known to be equivalent to Einstein
gravity (with the same fundamental scale) plus a scalar field theory with the scalar field
\begin{eqnarray}
\label{phig} \phi= \overline{M}_{D}^{(D-2)/2} \sqrt{\frac{D-1}{D-2}} \log \left|\frac{\partial f}{\partial
R}\right|
\end{eqnarray}
in a frame accessible by the conformal transformation $g_{A B} \rightarrow (\partial f/\partial R) g_{A B}$
\cite{ct}. Therefore, generalized action densities of the form $f({\cal{R}})$ are equivalent to scalar-tensor
theories of gravity, and thus, matter species are expected to experience an additional interaction due to the
exchange of the scalar field $\phi$ \cite{brans-dicke}. This is the fundamental signature of $f({\cal{R}})$
gravity compared to Einstein gravity for which simply $f({\cal{R}})= {\cal{R}}$. (Though remains outside the
scope of this work, see the discussions of Lovelock higher-curvature terms in \cite{rizzo}.)

In this work we study how $f({\cal{R}})$ gravity influences interactions among brane matter and certain collider
processes to observe them. In Sec. 2 below we derive graviton propagator and describe how it interacts with brane
matter. Here we put special emphasis on virtual graviton exchange. In Sec. 3 we study a number of higher
dimensional operators which are sensitive to $f({\cal{R}})$ gravity effects. In Sec. 4 we briefly discuss some
further signatures of $f({\cal{R}})$ gravity concerning graviton production and decay as well as certain loop
observables on the brane. In Sec. 5 we conclude.

\section{Gravitational Interactions from $f({\cal{R}})$ Gravity}
We parameterize generalized gravity theory via the action
\begin{eqnarray}
\label{fr} S=\int d^{D}x\, \sqrt{-g} \left\{ - \frac{1}{2} \overline{M}_D^{D-2} f\left({\cal{R}}\right) +
{\cal{L}}_{matter}\left(g_{A B}, \psi\right)\right\}
\end{eqnarray}
where couplings to matter fields $\psi$ are identical to those in (\ref{ah}). The metric field obeys
\begin{eqnarray}
\label{eom} f^{\prime}\left({\cal{R}}\right) {\cal{R}}_{A B} - \frac{1}{2} f\left({\cal{R}}\right) g_{A B} +
\left(g_{A B}\Box - \nabla_{A} \nabla_{B} \right)f^{\prime}\left({\cal{R}}\right) = - \frac{{\cal{T}}_{A
B}}{\overline{M}_{D}^{D -2}}
\end{eqnarray}
where prime denotes differentiation with respect to ${\cal{R}}$, and
\begin{eqnarray}
\label{tab} {\cal{T}}_{A B} = - \frac{2}{\sqrt{-g}} \frac{\delta \left(\sqrt{-g}
{\cal{L}}_{matter}\right)}{\delta g^{A B}} = \delta^{\delta}(\vec{y}) \delta_{A}^{\mu} \delta_{B}^{\nu} T_{\mu
\nu}(z)
\end{eqnarray}
is the stress tensor of the brane matter where $y_{i}$ and $z_{\mu}$ stand, respectively, for coordinates in
extra space and on the brane. The second equality here reflects the fact that entire energy and momentum are
localized on the brane. Clearly, energy-momentum flow has to be conserved $\nabla^{A} {\cal{T}}_{A B} =0$, and
this is guaranteed to happen provided that $\nabla^{\mu} T_{\mu \nu} = 0$.

Obviously, the equations of motion (\ref{eom}) reduce to Einstein equations when $f({\cal{R}})= {\cal{R}}$. In
general, for analyzing dynamics of small oscillations about a background geometry, $g_{A B} = g^0_{A B}$ with
curvature scalar ${\cal{R}}_{0}$, $f({\cal{R}})$ must be regular at ${\cal{R}}= {\cal{R}}_{0}$. In particular, as
suggested by (\ref{eom}), $f({\cal{R}})$ must be regular at the origin and $f(0)$ must vanish ($i.e.$ bulk
cosmological constant must vanish) for $f({\cal{R}})$ to admit a flat background geometry.

For determining how higher curvature gravity influences interactions among the brane matter, it is necessary to
determine the propagating modes which couple to the matter stress tensor. This requires expansion of the action
density in (\ref{fr}) by using (\ref{metric}) up to the desired order in $h_{A B}$. The zeroth order term
vanishes by $f(0) = 0$ constraint whose necessity was mentioned above. The terms first order in $h_{A B}$ vanish
by equations of motion. The quadratic part, on the other hand, turns out to be
\begin{eqnarray}
\label{sh} S_{h} = \int d^{D} x \left[\frac{1}{2} h_{A B}(x) {\cal{O}}^{A B C D}(x) h_{C D}(x) -
\frac{1}{\overline{M}_{D}^{(D-2)/2}} h_{A B}(x) {\cal{T}}(x)^{A B}\right]
\end{eqnarray}
such that propagator of $h_{A B}(x)$, defined via the relation
\begin{eqnarray}
{\cal{O}}_{A B C D}(x) {\cal{D}}^{C D E F}(x, x^{\prime}) = \frac{1}{2} \delta^{D}(x-x^{\prime})
\left(\delta^{E}_{A} \delta^{F}_{B} + \delta^{E}_{B} \delta^{F}_{A}\right)\,,
\end{eqnarray}
takes the form
\begin{eqnarray}
\label{prop} - i {\cal{D}}^{A B C D}(p^2) &=& - \left(\frac{f^{\prime}(0)+ 2 f^{\prime\prime}(0) p^{2}}{ (D-2)
f^{\prime}(0) + 2 (D-1)
f^{\prime\prime}(0) p^2}\right) \frac{1}{f^{\prime}(0) p^2} \eta^{A B} \eta^{C D}\nonumber\\
 &+& \frac{1}{2 f^{\prime}(0) p^2} \left(\eta^{A C} \eta^{B D} + \eta^{A D} \eta^{B C}\right)\nonumber\\
 &+& \frac{2 f^{\prime\prime}(0) p^{2}}{ (D-2) f^{\prime}(0) + 2 (D-1) f^{\prime\prime}(0)
p^2} \frac{1}{f^{\prime}(0) p^4} \left(\eta^{C D} p^A p^B + \eta^{A B} p^C
p^D\right)\nonumber\\
&+& \frac{(\xi -1)}{2 f^{\prime}(0) p^4} \left(\eta^{B D} p^A p^C + \eta^{D A} p^C p^B + \eta^{A C} p^B p^D +
\eta^{C B} p^D p^A\right)\nonumber\\
&+& \frac{2 (D -2) f^{\prime\prime}(0) p^{2}}{ (D-2) f^{\prime}(0) + 2 (D-1) f^{\prime\prime}(0) p^2}
\frac{1}{f^{\prime}(0) p^6} p^A p^B p^C p^D
\end{eqnarray}
in momentum space. It is clear that
\begin{eqnarray}
f^{\prime}(0) > 0
\end{eqnarray}
as otherwise all graviton modes become ghost. Therefore, if one is to prevent ghosty modes participating in
physical processes it is necessary to keep $f^{\prime}(0)$ positive definite. The parameter $\xi$ in (\ref{prop})
arises from the gauge fixing term
\begin{eqnarray}
{\cal{L}}_{g} = \frac{f^{\prime}(0)}{\xi} \eta^{A C}\left(\partial^{B} h_{A B} - \frac{1}{2} \partial_{A}
h_B^B\right) \left(\partial^{D} h_{C D} - \frac{1}{2} \partial_{C} h_D^D\right)
\end{eqnarray}
added to the $h_{A B}$ action density in (\ref{sh}). Here, $f^{\prime}(0)$ is introduced to match the terms
generated by ${\cal{L}}_{g}$ with the ones in (\ref{sh}). The propagator (\ref{prop}) depends explicitly on the
second derivative of $f({\cal{R}})$ evaluated at the origin, and it correctly reduces to the graviton propagator
in Einstein gravity \cite{giudice,han} when $f^{\prime\prime}(0)=0$ and $f^{\prime}(0) = 1$. The specific choice
$\xi =1$ corresponds to de Donder gauge frequently employed in quantum gravity. Obviously, one can probe
$f({\cal{R}})$ with higher and higher precision by computing higher and higher order $h_{A B}$ correlators.
Indeed, for probing $f^{\prime\prime\prime\prime}(0)$, for instance, it is necessary to expand the action density
in (\ref{fr}) up to quartic order so as to compute the requisite four-point function. Rather generically, higher
the order of correlators higher the dimensions of the operators they induce. The propagator (\ref{prop}) induces
a dimension-8 operator via graviton exchange between two matter stress tensors \cite{giudice,giudice2}. On the
other hand, four-point function induces a dimension-16 operator via graviton exchange among four matter stress
tensors.

The scattering processes which proceed with graviton exchange do exhibit new features as one switch from ADD
setup to $f({\cal{R}})$ gravity. Indeed, single graviton exchange influences various processes including
$2\rightarrow 2$ scatterings, particle self-energies, box diagrams and as such. The tree level processes are
sensitive to virtual states associated with the propagation of graviton in the bulk. On the other hand, loop
level processes involve particle virtualities both on the brane and in the bulk. In this sense, tree level
processes offer some degree of simplicity and clarity for disentangling the graviton contribution (see
\cite{giudice2} for a through analysis of the virtual graviton exchange effects) from those of the SM states.
Hence, in the following, we will restrict our discussions exclusively to tree level processes.

By imposing compactness of the extra space and taking its shape to be a torus as in the ADD mechanism one finds
\begin{eqnarray}
{\cal{T}}_{A B}(x) = \sum_{n_1=-\infty}^{+\infty}\cdots \sum_{n_{\delta}=-\infty}^{+\infty} \int \frac{d^4
p}{(2\pi)^4} \frac{1}{\sqrt{V_{\delta}}} e^{- i \left(k\cdot z - \frac{\vec{n}\cdot\vec{y}}{R}\right)}
\delta_{A}^{\mu} \delta_{B}^{\nu} {T}_{\mu \nu}(k)
\end{eqnarray}
where $\left(n_1, \dots, n_{\delta}\right)$ is a $\delta$-tuple of integers. Given this Fourier decomposition of
the stress tensor, the amplitude for an on-brane system $a$ to make a transition into another on-brane system $b$
becomes
\begin{eqnarray}
\label{amp} {\cal{A}}(k^2) = \frac{1}{\overline{M}_{Pl}^2} \sum_{\vec{n}} T_{\mu \nu}^{(a)}(k)
{\cal{D}}^{\mu\nu\lambda\rho}\left(k^2 - \frac{\vec{n}\cdot\vec{n}}{R^2}\right) T_{\lambda \rho}^{(b)}(k)
\end{eqnarray}
where use has been made of (\ref{planck}) in obtaining $1/\overline{M}_{Pl}^2$ factor in front. Though we are
dealing with a tree-level process the amplitude involves a summation over all Kaluza-Klein levels due to the fact
that these states are inherently virtual because of their propagation off the brane. Conservation of energy and
momentum implies that only the first two terms in the propagator (\ref{prop}) contributes to (\ref{amp}).
Therefore, after performing summation the transition amplitude (\ref{amp}) takes the form
\begin{eqnarray}
\label{as} {\cal{A}}(k^2)&=& \frac{S_{\delta-1}}{(2 \pi)^{\delta}}\, \frac{1}{\overline{M}_D^4 f^{\prime}(0)}
\left(\frac{\Lambda}{\overline{M}_{D}}\right)^{\delta-2} R\left(\frac{\Lambda}{\sqrt{k^2}}\right)
\left(T^{(a)}_{\mu \nu} T^{(b)\, \mu \nu} - \frac{1}{\delta +2} T_{\mu}^{(a)\,\mu}
T_{\nu}^{(b)\,\nu}\right)\nonumber\\
&+& \frac{(\delta+4)}{2(\delta+2)(\delta+3)}\, \frac{S_{\delta-1}}{(2 \pi)^\delta}\, \frac{1}{\overline{M}_D^4
f^{\prime}(0)} \left(\frac{\Lambda}{\overline{M}_{D}}\right)^{{\delta}-2}
R\left(\frac{\Lambda}{\sqrt{\tilde{k}^2}}\right) T_{\mu}^{(a)\,\mu} T_{\nu}^{(b)\,\nu}
\end{eqnarray}
which exhibits a huge enhancement ${\cal{O}}\left(\overline{M}_{Pl}^2/\overline{M}_D^2\right)$ compared to
(\ref{amp}) due to the contributions of finely-spaced Kaluze-Klein levels \cite{idea1}. Here $S_{\delta-1}= (2
\pi^{\delta/2})/\Gamma(\delta/2)$ is the surface area of $\delta$-dimensional unit sphere, $\tilde{k}^2 = k^2 -
m_{\phi}^2$ (see below eq. (18) for definitions), and $\Lambda$ (which is expected to be
${\cal{O}}\left(\overline{M}_D\right)$ since above $\overline{M}_D$ underlying quantum theory of gravity
completes the classical treatment pursued here) is the ultraviolet cutoff needed to tame divergent summation over
Kaluza-Klein levels. In fact, ${\cal{A}}(k^2)$ exhibits a strong dependence on $\Lambda$, as suggested by (see
also series expressions of $R\left({\Lambda}/{\sqrt{k^2}}\right)$ derived in \cite{giudice,han})
\begin{eqnarray}
\label{time} R\left(\frac{\Lambda}{\sqrt{k^2}}\right) =  - i \frac{\pi}{2}
\left(\frac{k^2}{\Lambda^2}\right)^{\frac{\delta}{2}-1} + \frac{\pi}{2}
\left(\frac{k^2}{\Lambda^2}\right)^{\frac{\delta}{2}-1} \cot \frac{\pi \delta}{2} - \frac{1}{\delta-2}\,\,
{}_2F_{1}\left(1, 1-\frac{\delta}{2}, 2 - \frac{\delta}{2},\frac{k^2}{\Lambda^2}\right)
\end{eqnarray}
for $0\leq k^2 \leq \Lambda^2$, and
\begin{eqnarray}
\label{space}
 R\left(\frac{\Lambda}{\sqrt{k^2}}\right) = \frac{1}{\delta} \frac{\Lambda^2}{k^2}\,\, {}_2F_{1}\left(1, \frac{\delta}{2}, 1
+ \frac{\delta}{2},\frac{\Lambda^2}{k^2}\right)
\end{eqnarray}
for  $k^2 <0$ or $k^2>\Lambda^2$. The imaginary part of $R$, relevant for the timelike propagator (\ref{time}),
is generated by exchange of on-shell gravitons $i.e.$ those Kaluza-Klein levels satisfying
$k^2=\vec{n}\cdot\vec{n}/R^2$. On the other hand, its real part follows from exchange of off-shell gravitons. For
spacelike propagator,  the scattering amplitude (\ref{space}) is real since in this channel Kaluza-Klein levels
cannot come on shell.

The first line of ${\cal{A}}({k^2})$ in (\ref{as}), except for the overall $1/f^{\prime}(0)$ factor in front, is
identical to the single graviton exchange amplitude computed within the ADD setup \cite{giudice,han}. In fact,
operators $T^{(a)}_{\mu \nu} T^{(b)\, \mu\nu}$ and $T_{\mu}^{(a)\,\mu} T_{\nu}^{(b)\,\nu}$ are collectively
induced by exchange of $J=2$ and $J=0$ modes of gravity waves $h_{A B}$ \cite{giudice,han}. The second line at
right-hand side, on the other hand, is a completely new contribution not found in ADD setup. The structure of the
induced interaction, $T_{\mu}^{(a)\,\mu} T_{\nu}^{(b)\,\nu}$, implies that it is induced by exchange of a scalar
field, different than the graviscalar which induces the same type operator in the first line of (\ref{as}). The
sources of this additional interaction is nothing but the scalar field $\phi$ defined in (\ref{phig}). Therefore,
the main novelty in ${\cal{A}}(k^2)$ lies in the second line at right-hand side of (\ref{as}) which is recognized
to be generated by the exchange of a scalar field with non-vanishing bare mass-squared
\begin{eqnarray}
\label{mphi} m_{\phi}^2 = -\frac{\delta+2}{2 (\delta+3)}\, \frac{f^{\prime}(0)}{f^{\prime\prime}(0)}
\end{eqnarray}
so that $\tilde{k^2} = k^2 - m_{\phi}^2$ in (\ref{as}). The nature of the scalar field $\phi$ depends on sign of
$f^{\prime\prime}(0)$: $\phi$ is a real scalar for $f^{\prime\prime}(0)< 0 $ and a tachyon for
$f^{\prime\prime}(0)> 0$. Moreover, when $f^{\prime\prime}(0) = 0$  it is clear that $f({\cal{R}})$ gravity
remains Einsteinian up to ${\cal{O}}({\cal{R}}^3)$ and this reflects itself by decoupling of $\phi$ from
propagator (\ref{prop}) and transition amplitude (\ref{amp}) since now $\phi$ is an infinitely massive scalar. On
the other hand, when $f^{\prime\prime}({\cal{R}})$ is singular at the origin the bare mass of $\phi$ vanishes and
thus ${\cal{A}}(k^2)$ simplifies to the first line of (\ref{as}) such that coefficient of $T_{\mu}^{(a)\,\mu}
T_{\nu}^{(b)\,\nu}$ changes from $-1/(\delta+2)$ to $-1/(2(\delta+3))$. A tachyonic scalar, $m_{\phi}^2 < 0$,
decouples from the transition amplitude (\ref{as}) as $k^2-m_{\phi}^2 \rightarrow \infty$. This can be seen from
the asymptotic behavior of (\ref{time}) by noting that on-shell graviton graviton exchange is shut off for
$k^2-m_{\phi}^2\geq \Lambda^2$. Similarly, a true scalar, $m_{\phi}^2 > 0$, also  decouples from the transition
amplitude (\ref{as}) when $k^2-m_{\phi}^2 \rightarrow -\infty$ as suggested by the asymptotic behavior
(\ref{space}). In the next section we will study higher dimensional operators induced by $f({\cal{R}})$ gravity
and their collider signatures.

\section{Higher Dimensional Operators from $f({\cal{R}})$ Gravity}
The impact of $f({\cal{R}})$ gravity on the transition amplitude (\ref{as}) is restricted to occur via the
dimension-8 operator $T_{\mu}^{(a)\,\mu} T_{\nu}^{(b)\,\nu}$. This operator involves traces of the stress tensors
of both systems $a$ and $b$. In general, trace of the energy momentum tensor, at tree level, is directly related
to the sources of conformal breaking in the system \cite{conformal}. It may be instructive to determine stress
tensors and their traces for fundamental fields. The energy and momentum of a massive vector field $A_{\mu}$ is
contained in the conserved stress tensor
\begin{eqnarray}
\label{j=1} T^{(J=1)}_{\mu \nu} = \eta_{\mu \nu} \left( \frac{1}{4} F^{\lambda \rho} F_{\lambda \rho} -
\frac{1}{2} M_{A}^2 A_{\lambda} A^{\lambda}\right) - \left(F_{\mu}^{\rho} F_{\nu \rho} - M_{A}^2 A_{\mu}
A_{\nu}\right)
\end{eqnarray}
whose trace
\begin{eqnarray}
\label{tracevector} T^{(J=1)\, \mu}_{\mu} = - M_{A}^{2} A_{\mu} A^{\mu}
\end{eqnarray}
demonstrates that vector boson mass breaks conformal invariance explicitly. On the other hand, conserved
energy-momentum tensor for a massive fermion reads as
\begin{eqnarray}
\label{j=1/2} T^{(J=1/2)}_{\mu\nu}&=& - \eta_{\mu \nu} \left( \overline{\psi} i \slashchar{\partial} \psi -
m_{\psi} \overline{\psi} \psi\right) + \frac{i}{2} \overline{\psi}\left( \gamma_{\mu} \partial_{\nu} +
\gamma_{\nu}
\partial_{\mu}\right) \psi\nonumber\\ &+& \frac{1}{4} \left[ 2 \eta_{\mu \nu} \partial^{\lambda}
\left(\overline{\psi} i \gamma_{\lambda} \psi\right) -
\partial_{\mu} \left(\overline{\psi} i \gamma_{\nu} \psi\right) - \partial_{\nu}
\left(\overline{\psi} i \gamma_{\mu} \psi\right) \right]
\end{eqnarray}
whose trace
\begin{eqnarray}
\label{tracefermion} T^{(J=1/2)\, \mu}_{\mu} = m_{\psi} \overline{\psi} \psi
\end{eqnarray}
shows that fermion mass breaks conformal invariance explicitly. In contrast to vector fields and spinors, trace
of the stress tensor for a scalar field is not directly related to its mass term. In fact, $T^{(J=0)\,
\mu}_{\mu}$ is nonzero even for a massless scalar. For a scalar field $\Phi$ to have $T^{(J=0)\, \mu}_{\mu}$ to
be proportional to its mass term it is necessary to introduce gauging $\Box \Phi \rightarrow \left(\Box - \zeta_c
{\cal{R}}\right) \Phi$ with `gauge coupling' $\zeta_c = (D-2)/(4(D-1))$ \cite{conf}. The curvature scalar serves
as the gauge field of local scale invariance. This gauging gives rise to additional terms in the stress tensor of
$\Phi$, and they do not vanish even in the flat limit. More explicitly, for a massive complex scalar with quartic
coupling the stress tensor reads as
\begin{eqnarray}
\label{j=0} T^{(J=0)}_{\mu \nu} &=& -\eta_{\mu \nu} \left[ \partial^{\rho} \Phi^{\dagger} \partial_{\rho} \Phi -
M_{\Phi}^2 \Phi^{\dagger} \Phi - \lambda \left(\Phi^{\dagger} \Phi\right)^2 \right] + \partial_{\mu}
\Phi^{\dagger}
\partial_{\nu} \Phi +
\partial_{\nu} \Phi^{\dagger}
\partial_{\mu} \Phi\nonumber\\
&+& 2 \zeta \left( \eta_{\mu \nu} \Box - \partial_{\mu} \partial_{\nu}\right) \Phi^{\dagger} \Phi
\end{eqnarray}
whose trace
\begin{eqnarray}
\label{tracescalar} T^{(J=0)\, \mu}_{\mu} = - 2 (1 - 6 \zeta) \left[ \partial^{\rho}\Phi^{\dagger}\partial_{\rho}
\Phi - \lambda \left(\Phi^{\dagger} \Phi\right)^2\right] + 4 \left(1 - 3 \zeta\right) M_{\Phi}^{2} \Phi^{\dagger}
\Phi
\end{eqnarray}
reduces to $T^{(J=0)\, \mu}_{\mu} = 2 M_{\Phi}^{2} \Phi^{\dagger} \Phi$ for $\zeta=\zeta_c\equiv 1/6$, as
desired. For $\zeta \neq \zeta_c$, say $\zeta =0$, $T^{(J=0)\, \mu}_{\mu}$ involves kinetic term,
self-interaction potential $\lambda \left(\Phi^{\dagger} \Phi\right)^2$ as well as mass term of the scalar field.
The terms proportional to $\zeta$ in (\ref{j=0}) might be regarded as either following from coupling of $\Phi$ to
curvature scalar as discussed above, or as a field-theoretic technicality to improve properties of the dilatation
current \cite{dilatation}.

The stress tensor traces (\ref{tracevector}), (\ref{tracefermion}) and (\ref{tracescalar}) with $\zeta=1/6$ show
that effects of graviscalar exchange (the operator $T^{(a)\, \mu}_{\mu} T^{(b)\, \nu}_{\nu}$ in the first line of
(\ref{as})) and $f({\cal{R}})$ gravity (the operator in the second line of (\ref{as})) can show up only in those
scattering processes which involve massive brane matter at their initial and final states. Their phenomenological
viability depends on how heavy the brane states compared to $\overline{M}_{D}$. For instance, high-energy
processes initiated by $e^+ e^-$ annihilation or $\gamma \gamma$ scattering or $p \overline{p}$ annihilation
cannot probe the operator $T^{(a)\, \mu}_{\mu} T^{(b)\, \nu}_{\nu}$ in (\ref{as}). On the other hand, scattering
processes which involve heavy fermions ($e.g.$ bottom and top quarks, muon and tau lepton), weak bosons
$W^{\pm}$, $Z$, and Higgs boson $h$ are particularly useful for probing the gravitational effects. Each of these
processes provides an arena for probing effects of scalar graviton exchange in general, and $f({\cal{R}})$
gravity effects in particular. It might be instructive to depict explicitly how ${\cal{A}}(k^2)$ differs from
that computed within the ADD setup by a number of specific scattering processes.

Concerning $2\rightarrow 2$ scattering of weak bosons one can consider, for instance, the process
$Z_{\alpha}(p_1) Z_{\beta}(p_2) \rightarrow Z_{\gamma}(k_1) Z_{\lambda}(k_2)$ which is described by the amplitude
\begin{eqnarray}
\label{azz} {\cal{A}}_{Z Z \rightarrow Z Z}(k^2) &=& {\cal{A}}_{SM}(k^2)+ \frac{1}{f^{\prime}(0)} {\cal{A}}_{ADD}(k^2)\nonumber\\
&+& \frac{(\delta+4)}{2 (\delta+2)(\delta+3)}\, \frac{S_{\delta-1}}{(2 \pi)^\delta}\,
\frac{M_Z^4}{\overline{M}_D^4
f^{\prime}(0)} \left(\frac{\Lambda}{\overline{M}_{D}}\right)^{{\delta}-2} \nonumber\\
&\times& \left\{ R\left(\frac{\Lambda}{\sqrt{\tilde{s}}}\right) \eta_{\alpha \beta} \eta_{\gamma \lambda}+
R\left(\frac{\Lambda}{\sqrt{\tilde{t}}}\right) \eta_{\alpha \gamma} \eta_{\beta \lambda} +
R\left(\frac{\Lambda}{\sqrt{\tilde{u}}}\right) \eta_{\alpha \lambda} \eta_{\beta \gamma} \right\}\nonumber\\
&\times& \epsilon_Z^{\alpha}(p_1) \epsilon^{\beta}_{Z}(p_2) \epsilon_Z^{\star\, \gamma}(k_1)
\epsilon_{Z}^{\star\, \lambda}(k_2)
\end{eqnarray}
after using (\ref{j=1}) in (\ref{as}). In this expression, $s=(p_1 +p_2)^2=(k_1+k_2)^2$, $t=(k_1-p_1)^2$ and
$u=(k_2-p_1)^2=4 M_Z^2 - s - t$ are Mandelstam variables, and $\epsilon^{\mu}_{Z}$ stands for the polarization
vector of $Z$ boson. The amplitudes ${\cal{A}}_{SM}(k^2)$ and ${\cal{A}}_{ADD}(k^2)$ can be found in \cite{vvvv}.
Obviously, $f({\cal{R}})$ gravity effects get pronounced when $M_D$ lies close to $M_Z$. Clearly, $\sigma\left(Z
Z \rightarrow Z Z\right)$ feels $f({\cal{R}})$ gravity via square of the third term in (\ref{azz}) and its
interference with SM and ADD contributions.

The fermion scattering $\psi_{1}(p_1) \psi_{1}(p_2) \rightarrow \psi_{2}(k_1) \psi_{2}(k_2)$ is described by
\begin{eqnarray}
\label{aff} {\cal{A}}_{\psi_1 \psi_1 \rightarrow \psi_2 \psi_2}(k^2) &=& {\cal{A}}_{SM}(k^2)+ \frac{1}{f^{\prime}(0)}
{\cal{A}}_{ADD}(k^2)\nonumber\\
&+& \frac{(\delta+4)}{2 (\delta+2)(\delta+3)}\, \frac{S_{\delta-1}}{(2 \pi)^\delta}\, \frac{m_{\psi_1}
m_{\psi_2}}{\overline{M}_D^4
f^{\prime}(0)} \left(\frac{\Lambda}{\overline{M}_{D}}\right)^{{\delta}-2} \nonumber\\
&\times& R\left(\frac{\Lambda}{\sqrt{\tilde{s}}}\right) \overline{\psi}_1(p_1) \psi_1(p_2) \overline{\psi}_2(k_1)
\psi_2(k_2)
\end{eqnarray}
after using (\ref{j=1/2}) in (\ref{as}). If $\psi_1$ and $\psi_2$ are identical fermions then $t$ and $u$ channel
contributions must also be included. The SM and ADD pieces in this amplitude can be found in \cite{giudice,han}.
The heavy fermion scatterings ($e.g.$ $t t \rightarrow t t$, $b b \rightarrow t t $, $\tau \tau \rightarrow t t$)
are potential processes for highlighting effects of $f({\cal{R}})$ gravity.

\jfig{boncuk1}{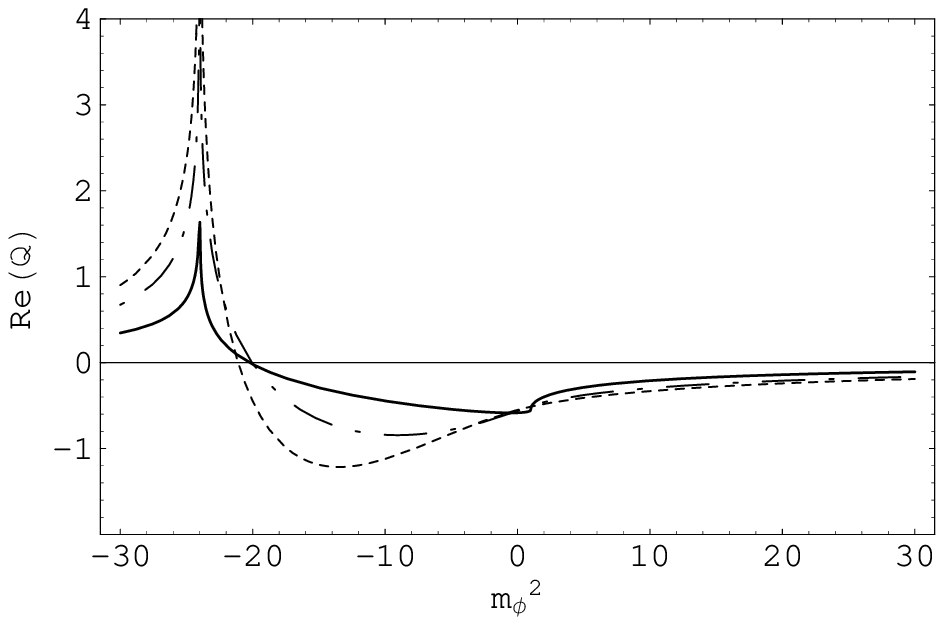}{ The dependence of $\mbox{Re}\left[{{Q}}(k^2)\right]$ on $m_{\phi}^2$ for $k^2= (1\,
{\rm TeV})^{2}$, $\Lambda = \overline{M}_{D} = 5\, {\rm TeV}$, and  $\delta=3$ (solid curve), $\delta=5$
(dot-dashed curve) and $\delta=7$ (short-dashed curve). We vary $m_{\phi}^2$ from $- (30\, {\rm TeV})^2$ up to $+
(30\, {\rm TeV})^2$.}

\jfig{boncuk2}{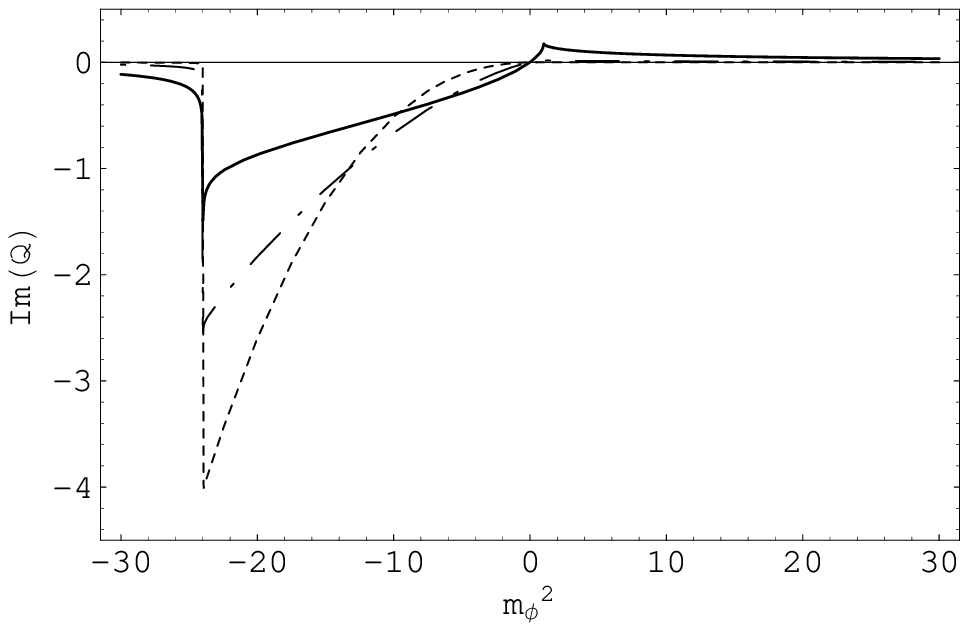}{ The same as in Fig. \ref{boncuk1} but for $\mbox{Im}\left[{{Q}}(k^2)\right]$.}

The $2\rightarrow 2$ scattering of Higgs bosons provides another interesting channel to probe $f({\cal{R}})$
gravity effects. Indeed, after expanding (\ref{j=0}) around the electroweak vacuum $\Phi = (v+h, 0)/\sqrt{2}$
with $v\simeq 246\, {\rm GeV}$, the amplitude for $h(p_1) h(p_2) \rightarrow h(k_1) h(k_2)$ scattering takes the
form
\begin{eqnarray}
\label{ahh} {\cal{A}}_{h h \rightarrow h h}(k^2) &=& {\cal{A}}_{SM}(k^2)+ \frac{1}{f^{\prime}(0)}
{\cal{A}}_{ADD}(k^2)\nonumber\\
&+& \frac{(\delta+4)}{8 (\delta+2)(\delta+3)}\, \frac{S_{\delta-1}}{(2 \pi)^\delta}\,
\frac{m_h^4}{\overline{M}_D^4 f^{\prime}(0)}
\left(\frac{\Lambda}{\overline{M}_{D}}\right)^{{\delta}-2}\nonumber\\ &\times& \left\{
R\left(\frac{\Lambda}{\sqrt{\tilde{s}}}\right)+
R\left(\frac{\Lambda}{\sqrt{\tilde{t}}}\right)+R\left(\frac{\Lambda}{\sqrt{\tilde{u}}}\right)\right\}
\end{eqnarray}
where $m_h^2 = -2 M_{\Phi}^2$ is the Higgs boson mass-squared. It is clear that size of $f({\cal{R}})$ gravity
effects depends crucially on how close $M_D$ is to $m_h$. Calculations \cite{xghe} within ADD setup show that
graviton exchange can have significant impact on $h(p_1) h(p_2) \rightarrow h(k_1) h(k_2)$, and thus, resulting
deviation from the SM expectation might be of observable size.

The $2 \rightarrow 2$ scattering processes mentioned above illustrate how $f({\cal{R}})$ gravity influences
certain observables to be measured in collider experiments. Beyond these, there are, of course various
observables which can sense $f({\cal{R}})$ gravity. For instance, $h Z Z$ coupling, which is crucial for Higgs
boson search via Bjorken process, gets also modified by graviton exchange \cite{hzz} via $T^{(J=0)}_{\mu \nu}\,
T^{(J=1)}_{\lambda \rho}$ correlator. The discussions above show that, independent of what brane matter species
are taking part in a specific process, entire novelty brought about by $f({\cal{R}})$ gravity is contained in the
second line of (\ref{as}), and thus, it proves useful to carry out a comparative analysis of this contribution
with the same structure present in the ADD setup, for completeness. In fact, ratio of the coefficients of
$T_{\mu}^{(a)\,\mu} T_{\nu}^{(b)\,\nu}$ in (\ref{as})
\begin{eqnarray}
{{Q}}(k^2)=-\frac{(\delta+4)}{2(\delta+3)}
\frac{R\left(\frac{\Lambda}{\sqrt{\tilde{k}^2}}\right)}{R\left(\frac{\Lambda}{\sqrt{{k}^2}}\right)}
\end{eqnarray}
is a useful quantity for such a comparative analysis. For determining how finite $f^{\prime\prime}(0)$ influences
the scattering processes it suffices to determine $m_{\phi}^{2}$ dependence of ${{Q}}(k^2)$ for given values of
$k^{2}$, $\delta$ and $\Lambda \sim \overline{M}_D$. In accord with future collider searches, one can take, for
instance, $k^2 = (1\, {\rm TeV})^{2}$ and $\Lambda = \overline{M}_{D} = 5\, {\rm TeV}$, and examine
$m_{\phi}^{2}$ dependencies of $\mbox{Re}\left[{{Q}}(k^2)\right]$ and $\mbox{Im}\left[{{Q}}(k^2)\right]$
separately. In fact, depicted in Figs. \ref{boncuk1} and \ref{boncuk2} are, respectively, the variations of
$\mbox{Re}\left[{{Q}}(k^2)\right]$ and $\mbox{Im}\left[{{Q}}(k^2)\right]$ with $m_{\phi}^{2}$. In the figures,
$m_{\phi}^2$ varies from $- (30\, {\rm TeV})^2$ up to $+ (30\, {\rm TeV})^2$ for each number of extra dimensions
considered: $\delta=3$ (solid), $\delta=5$ (dot-dashed) and $\delta=7$ (short-dashed). As suggested by
(\ref{mphi}), positive and negative $m_{\phi}^2$ values in the figures correspond, respectively, to negative and
positive values of $f^{\prime\prime}(0)$ since $f^{\prime}(0)$ has already been restricted to take positive
values to prevent graviton becoming a ghost (see the propagator (\ref{prop}). On the other hand, if
${\cal{A}}(k^2)$ in (\ref{as}) exhibits a timelike ($\tilde{k}^2>0$) or spacelike ($\tilde{k}^2<0$) propagation
depends on how $k^2$ compares with $m_{\phi}^2$. With the values of parameters given above, the figures
illustrate cases where $k^2>0$ yet $\tilde{k}^2$ varies over a wide range of values comprising spacelike and
timelike behaviors as well as a heavy $\phi$ $i.e.$ $\left|m_{\phi}^{2}\right| \gg \Lambda^2$.

The overall behaviors of both figures suggest that $F({\cal{R}})$ gravity effects fade away for large
$\left|m_{\phi}^{2}\right|$, as expected. Both real and imaginary parts of ${{Q}}(k^2)$ exhibit a narrow peak at
$m_{\phi}^2 = - (24\, {\rm TeV})^{2}$ which corresponds to resonating of the transition amplitude by Kaluza-Klein
levels with mass-squared $=k^2-m_{\phi}^2=\Lambda^2$. From Fig. \ref{boncuk1} it is clear that
$\mbox{Re}\left[{{Q}}(k^2)\right]$ becomes significant at large $\delta$ and negative $m_{\phi}^2$. This is also
seen to hold for $\mbox{Im}\left[{{Q}}(k^2)\right]$ from Fig. \ref{boncuk2}. Obviously, $f({\cal{R}})$ gravity
predictions differ from ADD ones for moderate (with respect to scale $\Lambda$) negative $m_{\phi}^2$ or
equivalently for sufficiently small and positive $f^{\prime\prime}(0)$ (see eq.(\ref{mphi}) for details). Indeed,
for positive values of $m_{\phi}^{2}$ or equivalently for negative $f^{\prime\prime}(0)$ the strength of
$f({\cal{R}})$ gravity contribution remains significantly below the ADD one. The behaviors of
$\mbox{Re}\left[{{Q}}(k^2)\right]$ and $\mbox{Im}\left[{{Q}}(k^2)\right]$ illustrated in these figures remain
qualitatively similar for other $k^2$, $\Lambda$ or $\delta$ combinations. Therefore, one concludes that
generalization of ADD mechanism based on Einstein-Hilbert action density $\overline{M}_D^2 {\cal{R}}$ to
$\overline{M}_D^2 f\left({\cal{R}}\right)$ type higher curvature gravity theory can have observable effects on
high energy processes (for instance, the ones listed in (\ref{azz}), (\ref{aff}) and (\ref{ahh}) listed above)
when $f^{\prime\prime}(0)$ is positive and ${\cal{O}}\left(1/\overline{M}_D^{2}\right)$.

\section{ Yet More Signatures of $f\left({\cal{R}}\right)$ Gravity}
So far we have focussed mainly on higher dimensional operators induced by tree-level virtual graviton exchange.
Clearly, effects of higher dimensional gravity on brane matter are not restricted to such processes: graviton can
contribute to self-energies, effective vertices or box diagrams of brane matter; graviton can be emitted off the
brane matter; and graviton can decay into brane matter. In this section we will discuss such processes briefly
for illustrating how $f\left({\cal{R}}\right)$ gravity effects differ from those found in the ADD setup.

First of all, as suggested by (\ref{fr}), couplings of the gravity waves $h_{A B}(x)$ to brane matter are
identical in ADD and $f\left({\cal{R}}\right)$ gravity setups. Therefore, distinction between the two frameworks
rests mainly on the additional scalar field (\ref{phig}) imbrued in the $f\left({\cal{R}}\right)$ gravity
dynamics. Consequently, detection of $f\left({\cal{R}}\right)$ gravity effects requires scattering processes on
the brane to be sensitive to the new energy threshold $m_{\phi}$ not found in the ADD setup.

Let us consider first role of $f\left({\cal{R}}\right)$ gravity on brane-localized loops. The simplest of such
processes is the self-energy of a brane particle. One may consider, for instance, self-energy of the $Z$ boson
(or any of the massive SM fields mentioned in the last section). At the level of a single graviton exchange one
finds
\begin{eqnarray}
\label{self} -i \Pi\left(q\right) &=& -i \Pi_{SM}\left(q\right) - i\frac{1}{f^{\prime}(0)}
\Pi_{ADD}\left(q\right) -
i \Pi_{seagull}\left(q^2\right)\\
&-&\frac{M_Z^4}{\overline{M}_{Pl}^2 f^{\prime}(0)} \frac{\delta+4}{ (\delta+2) (\delta+3)} \sum_{\vec{n}} \int
\frac{d^4 k}{(2 \pi)^4} \frac{1}{k^2 - m_{\phi}^2 - \frac{\vec{n}\cdot\vec{n}}{R^2}+i\epsilon}\, \frac{1}{(q+k)^2
- M_{Z}^2+i \epsilon}\nonumber
\end{eqnarray}
where contribution of the four-point vertex that binds gravitons and $Z$ bosons is contained in the seagull
contribution. The summation and integration involved in this expression are difficult to evaluate analytically,
and therefore, one may eventually need to resort some numerical techniques \cite{han}. However, at least for
vanishing external momentum, one can show that $f\left({\cal{R}}\right)$ gravity contribution in the second line
of (\ref{self}) is diminished at large $\left|m_{\phi}\right|^2$, and is particularly pronounced when
$\left|m_{\phi}\right|^2 \sim M_Z^2$ and $m_{\phi}^2<0$. Therefore, when $f^{\prime\prime}(0) \sim 1/M_{Z}^2$ one
expects observable enhancements in the $Z$ boson self energy (see \cite{giudicex} and \cite{han} for analyses of
the Higgs boson self energy).

Having discussed effects of $f\left({\cal{R}}\right)$ gravity on brane-localized loops we now turn to an analysis
of production and decays of the graviton. In these processes graviton is a physical particle described by
asymptotically free states connected by the S-matrix elements. Therefore, the scalar field $\phi$ imbrued in
$f\left({\cal{R}}\right)$ dynamics must be endowed with a positive mass-squared for its decays and productions to
be observable. Consequently, $f\left({\cal{R}}\right)$ gravity effects on graviton production and decay exist
within $m_{\phi}^2 > 0$ domain. However, as suggested by Figs. \ref{boncuk1} and \ref{boncuk2},
$f\left({\cal{R}}\right)$ gravity contribution, the second line of (\ref{as}), stays significantly below the
corresponding contribution in ADD setup. This implies, in particular, that production and decay of $\phi$
graviton are suppressed relative to those of the $J=2$ and $J=0$ gravitons.

The above observation is confirmed by the fact that when the looping particles come on their mass shells, as
dictated by the optical theorem, the $Z$ boson self-energy (\ref{self}) above represents the Drell-Yan production
of graviton and $Z$ boson at lepton (via $e^+ e^- \rightarrow Z^{\star} \rightarrow graviton\; + Z$ annihilation)
or hadron (via $q \overline{q} \rightarrow Z^{\star} \rightarrow graviton\; + Z$ annihilation) colliders. The
main novelty brought about by $f\left({\cal{R}}\right)$ gravity  is the production of $\phi$ (in addition to
$J=0$ and $J=2$ gravitons) when the center-of-mass energy of the collider is sufficiently large $i.e.$ $s \geq
m_{\phi}^2 + M_{Z}^2$. This phenomenon reflects itself by a sudden change in the number of events (similar to
opening of $W^+ W^-$ channel at LEP experiments). The dominant contribution to graviton emission comes from
Kaluza-Klein levels in the vicinity of $R^2 (M_Z^2 - m_{\phi}^2)$. The emission of gravitons from the brane is
not restricted to such $2 \rightarrow 2$ processes, however. Indeed, massive brane-localized states can decay
into gravitons, including $\phi$ itself, and this reflects itself as an increase in their invisible widths (see,
for instance, \cite{giudicex} for a detailed discussion of the Higgs boson width).

There are, of course, inverse processes to graviton emission. Indeed, gravitons propagating in the bulk can decay
into brane matter when they land on the brane. The graviton decay channels can open only if their Kaluza-Klein
level is high enough \cite{han}. The only exception to this is the $\phi$ graviton which can decay into brane
matter even at zeroth Kaluza-Klein level provided that its mass, $m_{\phi}$, is larger than those of the daughter
particles. Detailed discussions of the production and decays of gravitons (as well as those of the right-handed
neutrinos propagating in the bulk \cite{gher}) in the framework of ADD mechanism can be found in \cite{idea1}.

This section is intended to provide a brief summary of what impact $f\left({\cal{R}}\right)$ gravity can have on
processes involving brane-loops, missing energy signals in brane matter scatterings, and population of brane via
the graviton decays. These processes are of great importance for both collider \cite{giudice,han} and
cosmological \cite{idea1,gher} purposes, and discussions provided in this section is far from being sufficient
for a proper description of what effects $f\left({\cal{R}}\right)$ gravity can leave on them. From this brief
analysis, combined with results of the previous section, one concludes that $f\left({\cal{R}}\right)$ gravity
effects on decays and emissions of graviton cannot compete with the ADD expectations.

\section{Conclusion}
In this work we have discussed a number of phenomenological implications of $f({\cal{R}})$ gravity in higher
dimensional spacetimes with large extra spatial dimensions. In Sec. 2 we have derived graviton propagator about
flat Minkowski background (which requires $f(0)$ $i.e.$ cosmological constant to vanish), and have determined how
it influences interactions among the brane matter. In Sec. 3 we have listed down a set of higher dimensional
operators which exhibit an enhanced sensitivity to $f({\cal{R}})$ gravity (compared to those operators involving
light fermions or massless gauge fields). Finally, in this section we have performed a comparative study of ADD
and $f({\cal{R}})$ gravity predictions and determined ranges of parameters where the latter dominates over the
former. The analysis therein suggests that $f({\cal{R}})$ gravity theories with finite and positive
$f^{\prime\prime}(0)$ induce potentially important effects testable at future collider studies. In Sec. 4 we have
discussed briefly how $f({\cal{R}})$ gravity influences loop processes on the brane as well as decays and
productions of gravitons.

The analysis in this work can be applied to various laboratory, astrophysical and cosmological observables (see
\cite{idea1} for a detailed discussion of major observables) for examining non-Einsteinian forms of general
relativity in higher dimensions. The discussions presented here are far from being complete in their coverage and
phenomenological investigations. The rule of thumb to be kept in mind is that higher curvature gravity influences
scatterings of massive (sufficiently heavy compared to the fundamental scale of gravity) brane matter.

\section{Acknowledgements}
 The work of D. A. D. was partially supported by Turkish Academy of Sciences through GEBIP grant,
 and by the Scientific and Technical Research Council of Turkey through project 104T503.

\end{document}